**Non-uniform Curvature and Anisotropic Deformation Control Wrinkling Patterns on Tori**

Xiaoxiao Zhang[1], Patrick T. Mather[2], Mark J. Bowick[3*], Teng Zhang[1†]

1. Department of Mechanical and Aerospace Engineering, Syracuse University, Syracuse, NY, 13244, USA

2. Department of Chemical Engineering, Bucknell University, Lewisburg, PA 17837 USA

3. Kavli Institute for Theoretical Physics, University of California, Santa Barbara, CA 93106, USA.

[*]bowick@kitp.ucsb.edu

[†]tzhang48@syr.edu

**Abstract**

We investigate wrinkling patterns in a tri-layer torus consisting of an expanding thin outer layer, an intermediate soft layer and an inner core with a tunable shear modulus, inspired by pattern formation in developmental biology, such as follicle pattern formation during the development of chicken embryos. We show from large-scale finite element simulations that hexagonal wrinkling patterns form for stiff cores whereas stripe wrinkling patterns develop for soft cores. Hexagons and stripes co-exist to form hybrid patterns for cores with intermediate stiffness. The governing mechanism for the pattern transition is that the stiffness of the inner core controls the degree to which the major radius of the torus expands – this has a greater effect on deformation in the long direction as compared to the short direction of the torus. This anisotropic deformation alters stress states in the outer layer which change from biaxial (preferred hexagons) to uniaxial (preferred stripes) compression as the core stiffness is reduced. As the outer layer continues to expand, stripe and hexagon patterns will evolve into zigzag and segmented labyrinth, respectively. Stripe wrinkles are observed to initiate at the inner surface of the torus while hexagon wrinkles start from the outer surface as a result of curvature-dependent stresses in the torus. We further discuss the effects of elasticities and geometries of the torus on the wrinkling patterns.

**1 Introduction**

Wrinkling patterns are ubiquitous in nature and engineering structures[1-16], from follicle pattern in the avian skin[6] to folded tissues[7-9], from stretchable electronics[10] to swelling gels[14-16]. To explore the fundamental principles of the formation and evolution of wrinkling patterns, layered soft materials are widely adopted in experiments and theories[17-23], such as a bi-layer structure with a relatively stiff film (coating) bonded to a relatively compliant substrate. Taking the flat bi-layer structure as an example, it has been well documented that the wrinkling patterns are determined by the stress states in the film and the ratio between the shear modulus of the film and substrate[17-19, 24-28]. Recent work has focused on the effect of curvature on the wrinkling patterns[28-41]. Hexagonal patterns, for example, were predicted to be more stable on curved substrates [28] and observed in elastomer spherical bi-layer structures[29-32]. Dimpled patterns were found to form crystalline-like structure with topological defects on spherical surface[31]. A recent study further showed that substrate curvature delays the critical strain for wrinkles on a cylinder[40]. Although non-uniform curvatures have been shown to play important roles in the spheroidal shapes of natural fruits and vegetables[41], there is still a general lack of understanding of the formation and evolution of wrinkling patterns on surfaces with curvature gradients.

In this paper, we study the wrinkling patterns on a torus because it is a simple model with variable curvatures and closely mimics the shape of an embryo during early development[42]. The torus contains three layers, including an outer film bonded to an intermediate layer as well as an inner core with tunable stiffness (Fig. 1). The materials in the three layers are modeled as a nearly incompressible neo-Hookean model with the same Poisson ratio ($v_0 = v_1 = v_2 = 0.475$). The outer film is 10 times stiffer than the intermediate layer (i.e., $\mu_0/\mu_1 = 10$) and undergoes isotropic volumetric expansion to drive surface wrinkling on the torus. The inner core controls the global deformation of the torus. This geometric confinement may be characterized by the elastic ratio $C = \mu_2/\mu_1$ between the shear moduli of the core and the intermediate layer.

This tri-layer torus serves as a minimal mechanical model of the skin morphologies in growing embryos, where the outer and intermediate layers can be seen as epidermis and dermis [6, 43-45], respectively. The stiff core can mimic the skeleton. In real biological structures, each component can have a different growth rate. The geometric confinement of the skeleton is not only determined by the modulus but also the relatively small growth rate compared to other components. To reduce the number of parameters in our model, we only let the outer layer expand and vary the modulus of the inner core to tune the geometric confinement. The elastic moduli of epidermis and dermis during the early embryo development have not been reported yet, although numerous studies have focused on the mechanical properties of mature skins [46-48]. Here, we choose $\mu_0/\mu_1 = 10$, which leads to the generation of stable hexagon patterns for a bi-layer mechanical structure subject to equal biaxial compression. It should also be noted that active tissue contraction in the dermis[6] is not explicitly modeled here and deserves further attention.

From fully three-dimensional finite element (FE) simulations, we observe the onset of a wrinkling pattern that transitions from stripes to hexagons as the elastic ratio $C$ is increased from 0.01 to 1000 and identify eight different wrinkling patterns, which can be categorized into three groups based on the patterns near the onset of wrinkles. In contrast, only one group, starting from the hexagon, is found on cylinders and spheres. By analyzing the stress states in the outer film in the tri-layer torus, we uncover the important roles of the non-uniform curvature and anisotropic deformation in determining the initiation and propagation of the wrinkling patterns. A phase diagram of the eight wrinkling patterns on the torus is established in terms of expansion of the outer film and elastic ratio $C$, which represents the driving force for wrinkles and confinement, respectively. The rich wrinkling patterns revealed in our study will provide rational guidelines for designing multifunctional structures with complicated surface morphologies and may also shed light on the developmental patterns found in animal hairs, feathers and scales.

## 2 Results

### 2.1 Pattern transition at different confinements

We first investigated the effect of the elastic ratio on the formation and evolution of wrinkling patterns on the torus. Only a quarter of the tri-layer torus was employed in this section thanks to the symmetry of the structure. All the simulations were performed with ABAQUS[49], the detailed information of which can be found in the supplementary materials. As we varied the elastic ratios (increasing level of constraint for the torus major radius) and the outer film expansion, we observed three typical wrinkling patterns, as shown in Fig.2. Note that the color scale corresponds to values of maximum principal logarithmic strain. For the smallest elastic ratio ($C = 1$), partial stripe patterns initiate at the inner surface of the torus at a critical expansion. As the expansion further increases, the stripe patterns propagate through the whole torus and eventually become zigzag patterns. For the largest elastic ratio ($C = 100$), featuring significant constraint of the torus major radius, partial hexagonal patterns occur at the outer surface first, propagate to the inner surface and then develop into segmented labyrinth patterns. For the intermediate elastic ratio ($C = 20$), stripe patterns first emerge and both stripe and hexagonal patterns coexist for a large expansion. This may provide a new way to make a Janus torus featuring different patterns at the inner and outer surfaces. At even larger expansion, both zigzag and segmented labyrinth patterns are observed.

Hexagon-like patterns were reported in studies of follicle patterns in avian skin and chicken skin samples cultured on gels [6, 43]. A similar spreading wave of the hexagon pattern was also found in feathers in avian skin, which was attributed to the integration of biological signaling and mechanochemical coupling [43]. In addition, it is interesting to note that the epidermal scales in reptiles often emerge first at the outer surface of the embryo[42]. Although the pattern formation in the reptilian embryo involves reaction-diffusion[50-53], growth[6] and mechanochemical coupling[45], the curvature effects on wrinkling patterns we have observed here are very likely to exist in these biological patterns by virtue of the significant curvatures present. This calls for future studies that directly connect to developmental mechanobiology systems.

**2.2 Phase diagram of the wrinkling patterns on torus**

We observed eight wrinkling patterns in three typical torus structures in the proceeding section, which include partial stripe, full stripe, partial hexagon, full hexagon, hybrid stripe-hexagon (H1), zigzag, segmented labyrinth, and hybrid zigzag-segmented labyrinth (H2). We ran more simulations to cover a wider range of the elastic ratio ($C = \mu_2/\mu_1 = 0.01 \sim 1000$) and constructed a phase diagram of these wrinkling patterns in terms of the elastic ratio and expansion value (Fig. 3). It should be noted that we identified different wrinkling "phases" by visual inspection. While this qualitative approach does not capture exact phase boundaries, it does reveal salient features of the phase diagram. Among these patterns, zigzag, segmented labyrinth and their hybrids are post-buckling patterns which evolve from the onset of wrinkling. We focused therefore on the patterns near the onset of wrinkles and cast them into three wrinkling regimes. We recognized that at lower $C$ (0.01~10), the first stable wrinkling pattern is dominated by stripes, which is defined as regime I. The stripe pattern later develops into a zigzag pattern during the post-buckling at large deformation as the expansion increases. When the elastic ratio $C$ exceeds 10, the stripe pattern first covers the full surface of the torus following which a hexagon pattern appears at the outer surface while stripes remain in the region near the inner surface, yielding a hybrid stripe-hexagonal pattern (H1). We denoted this as regime II. We found that the upper bound of the elastic ratio for this regime is 30. This hybrid stripe-hexagonal pattern transitions into hybrid zigzag-segmented labyrinth for large expansions. If we further increase the elastic ratio, we see a third regime ($C = \mu_2/\mu_1 = 30\sim1000$) where the first stable pattern is hexagonal. The hexagonal pattern in regime III first occurs at the outer surface and propagates to the inner surface at larger expansion and finally transitions into segmented labyrinth. It is interesting to notice that global buckling of a bi-layer hydrogel torus has been recently achieved by printing aqueous precursor in a yield stress fluid that preserves the toroid shape until polymerization[54]. This may provide a promising means to explore the phase diagram of wrinkling patterns predicted in our work.

**2.3 Stress states in the outer film**

It is well known that in the case of uniaxial compression, wrinkles usually exhibit stripe patterns (sinusoidal profiles)[21]. When subjected to equal biaxial compressive stresses, a thin film bonded on a substrate can buckle into different patterns such as checkerboard, herringbone, hexagonal, and labyrinthine[25-28]. To gain insights into the pattern evolution on tori, we investigated the distribution of the compressive stresses in the poloidal ($\sigma_{\Phi\Phi}$) and toroidal ($\sigma_{\theta\theta}$) directions. Here, toroidal and poloidal directions represent the long and short circular paths around the torus (Fig. 1a), respectively. We noticed that both poloidal ($\sigma_{\Phi\Phi}$) and toroidal ($\sigma_{\theta\theta}$) stresses are negative, so we only focused on their magnitude in the following analysis. Prior to wrinkling, the stresses in the outer film are periodic functions of $\Phi$ for tri-layer tori with different elastic ratios, as shown in Fig. 4(a). Since the stresses inside the outer film are caused by constrained expansion, global deformation of the torus will relax the constraints and thus tend to release the stresses. It can be seen in Fig. 4(a) that both poloidal ($\sigma_{\Phi\Phi}$) and toroidal ($\sigma_{\theta\theta}$) stress decrease when reducing the elastic ratios because the global deformation increases as the elastic ratio decreases. The toroidal stress, however, drops much faster than the poloidal stress, indicating an anisotropic confinement. The stresses vary along the poloidal direction with maxima or minima at $\Phi = 0, 2\pi$ and $\Phi = \pi$, which correspond to the center of the outer and inner side of the torus, respectively. This is consistent with our observation that wrinkling patterns can initiate from the center of the outer or the inner surface. We analyzed therefore the stress components of these two specific locations (the center of the outer surface $\Phi = 0, 2\pi$ and the center of the inner surface $\Phi = \pi$) and computed the ratio between the poloidal and toroidal stresses $\sigma_{\Phi\Phi}/\sigma_{\theta\theta}$ (Fig. 4(b)). By comparing the three regimes in the wrinkling phase diagram in Figure 3, we found that, in regime I, $\sigma_{\Phi\Phi}/\sigma_{\theta\theta}$ is larger than 1.10 at both locations, which deviates from equal biaxial compressive states and thus explains the initiation and propagation of stripe patterns. In regime III, the stress ratios at both locations are lower than 1.04, close to the equal biaxial compressive stress state, which can lead to hexagonal patterns on curved structures. Regime II is more complicated, as demonstrated in more wrinkling patterns and the intermediate ranges of the stresses. It can be seen that the stress states tend more toward uniaxial compression at the inner surface and toward biaxial compression at the outer surface. This explains why the stripe pattern starts at the inner surface while the hexagonal pattern emerges at the outer surface later.

We next sought to understand the locations of pattern initiation in light of the stress distributions associated with non-uniform curvatures. In regime I, the value of $\sigma_{\Phi\Phi}$ determines the location of the pattern initiation because the stripes all orient along the toroidal direction. We have observed in Fig. 4(a) that $\sigma_{\Phi\Phi}$ is larger at the center of the inner surface than the outer surface, which causes the stripe pattern to originate at the inner surface. In regime III, both $\sigma_{\Phi\Phi}$ and $\sigma_{\theta\theta}$ have maximum values at the center of the outer surface, so the hexagonal patterns start there.

## 3 Discussion

### 3.1 No pattern transition on cylindrical and spherical structures

To check whether the transition between stripe and hexagon is unique to the torus due to its non-uniform curvature and anisotropic confinement, we ran simulations of a tri-layer structure of a cylinder and a sphere, with the same cross-sections in Fig. 1(b). These cases feature constant mean and Gaussian curvatures. In the simulations of cylinders, we applied symmetric boundary conditions to the two ends along the axial direction to mimic an infinitely long structure, which fully confines the global deformation. For spherical structures, we only simulated one-eighth of the structure due to symmetry and can observe isotropically global deformation. In other words, the sphere is under isotropic confinement. For all elastic ratios studied, we only obtained hexagon patterns and did not observe transitions to stripes in cylinders and spheres (Fig. 5). We further explored the radius of the cylinder and sphere as parameters and found that the critical wrinkling strains increased systematically with mean curvature (see Fig. S1, ESI). These findings agree well with a recent study on curvature-delayed growth of wrinkles on cylinders[40]. This means that wrinkles should appear earlier for lower mean curvatures for structures with uniform curvature. This is another difference from wrinkles on the torus in regime III, where hexagons initiate at regions of higher curvature.

### 3.2 Effect of elasticities and geometries on the pattern transitions on tori

So far, our discussion on the torus is based on a fixed modulus ratio between the outer and intermediate layers (i.e., $\mu_0/\mu_1 = 10$) and the same geometry (i.e., $R = 150$ and $a = 31$). We further explored a range of material and geometry properties in the simulations to test the generality of our findings. We first changed $\mu_0/\mu_1$ from 10 to 20 and confirmed that stripe patterns form at low confinement (i.e., $\mu_2/\mu_1 = 10$) and hexagonal patterns at high confinement (i.e., $\mu_2/\mu_1 = 100$) (see Fig. S2, ESI). We then ran simulations of a structure with $a = 41, R = 150$ and observed the similar pattern transition from low confinement to high confinement (see Fig. S3, ESI). It should be noted that labyrinth-like wrinkles are found on spheres with large radius-to-thickness ratio[37]; however, this cannot be observed in the current simulations because of prohibitively high computational demand. Nevertheless, we expect more wrinkling patterns can emerge if we further enlarge the parameter spaces of the modulus ratio $\mu_0/\mu_1$ and radius-to-thickness ratio $a/h$.

For these slender tori, we have observed that stripe patterns initiate at the inner surface and hexagonal patterns start at the outer surface of the torus. For more general tori, their mean curvature and Gaussian curvature can be expressed as,

$$H = \frac{R+2a\cos\Phi}{2a(R+a\cos\Phi)} \text{ and } \kappa = \frac{\cos\Phi}{a(R+a\cos\Phi)}. \tag{1}$$

It can be seen that the mean curvature can change sign for tori with ratio between $a$ and $R$ larger than 0.5. To test the effect of negative mean curvature on the wrinkling patterns, we compared the results from two simulations with $a = 41, R = 70$, and $a = 81, R = 150$. For these two cases, we fixed $h = 1$, $r_1/r_2 = 3$, and $\mu_2/\mu_1 = 10$. For the smaller torus ($a = 41, R = 70$), we observed a toroidally oriented stripe pattern at a low confinement and a hybrid pattern at high confinement (Fig. 6a). For the larger torus ($a = 81, R = 150$), we observed a hybrid pattern at a low confinement and the hexagon pattern at high confinement (Fig. 6b). For these simulated tori with larger aspect ratios (i.e., a/R > 1/2), we also found compressive stresses in the poloidal ($\sigma_{\Phi\Phi}$) and toroidal ($\sigma_{\theta\theta}$) directions, which vary similarly to the slender tori as we change the confinement. This indicates that the same mechanism governs the pattern transitions on tori with different aspect ratios. Post-buckling patterns are not shown here because we focused on the pattern transitions. As can be seen in Fig. 6, the values of the mean and Gaussian curvatures have big differences in these two structures, which may explain the formation of different patterns. To fully elucidate the coupling effects between the gradients and values of the curvatures, future investigation of wrinkling patterns on ellipsoids and surfaces with zero mean curvature (e.g., catenoids) or zero Gaussian curvature (e.g., cones) are needed. In addition, it is harder to globally deform the intermediate layer in the larger torus because the required force scales with the area of the cross section. This additional confinement may eliminate the formation of a fully developed stripe pattern.

**4 Conclusions**

We found eight wrinkling patterns in tri-layer tori and established a phase diagram to describe these patterns. The key control variables of the phase diagram are the film expansion and the anisotropic deformation, characterized by the elastic ratio between the inner core and intermediate layer. Stripe patterns emerge at weak confinements because the anisotropic deformation of the torus relaxes the stress along the toroidal direction and makes the compressive strain along the circular direction more dominant. At strong confinements (large $C$), hexagons appear due to nearly equal biaxial compressive stress in the outer film. Hybrid stripe-hexagon patterns happen at intermediate confinement levels. In addition, the non-uniform curvature leads to a stress gradient in the outer film, despite the uniform driving force (expansion of the outer film). This stress gradient causes stripes to initiate at the inner surface, and hexagons to initiate from the outer surface.

We further found that the size and shapes of the torus play roles in determining the wrinkling patterns. For some tori, only a subset of the wrinkling patterns discovered here are observed. A phase diagram of wrinkling patterns with higher dimensional order is needed for general tori, which will be investigated in future studies. It should also be noted that real biological structures are more complicated than the purely mechanical system we adopted here. Therefore, another important future research direction will seek to understand the coupling effects of the elasticity, chemical signaling, and curvature on various biological patterns during the early development of embryos.

**Conflicts of interest**

There are no conflicts of interests to declare.

**Acknowledgements**

TZ thanks Pedro Reis for valuable discussion. Simulations were performed at the Triton Shared Computing Cluster at the San Diego Supercomputer Center, the Comet cluster (Award no. TG-MSS170004 to TZ) in the Extreme Science and Engineering Discovery Environment (XSEDE) and the Academic Virtual Hosting Environment (AVHE) at Syracuse University. The research of MJB was supported by the National Science Foundation under Grant No. NSF PHY-1748958.

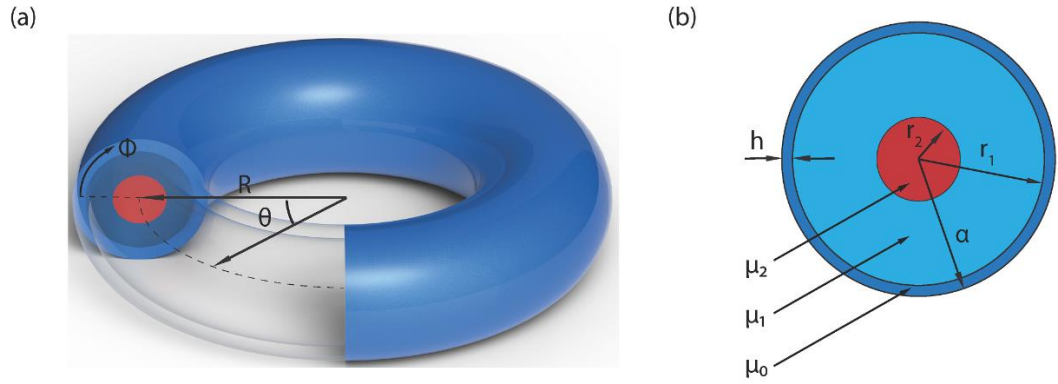

**Fig. 1.** (a) Schematic of the torus (major radius $R = 150$, minor radius $a = 31$). (b) Cross-section of the tri-layer torus with a solid film (dark blue, shear modulus $\mu_0$, thickness $h = 1$) adhered to a softer substrate (light blue, shear modulus $\mu_1$, radius $r_1 = 30$) and an inner core (red, shear modulus $\mu_2$, radius $r_2 = 10$). We normalize all the lengths with the thickness of the outer film ($h$) and all the elastic moduli as well as stress with the shear modulus of the intermediate layer ($\mu_1$).

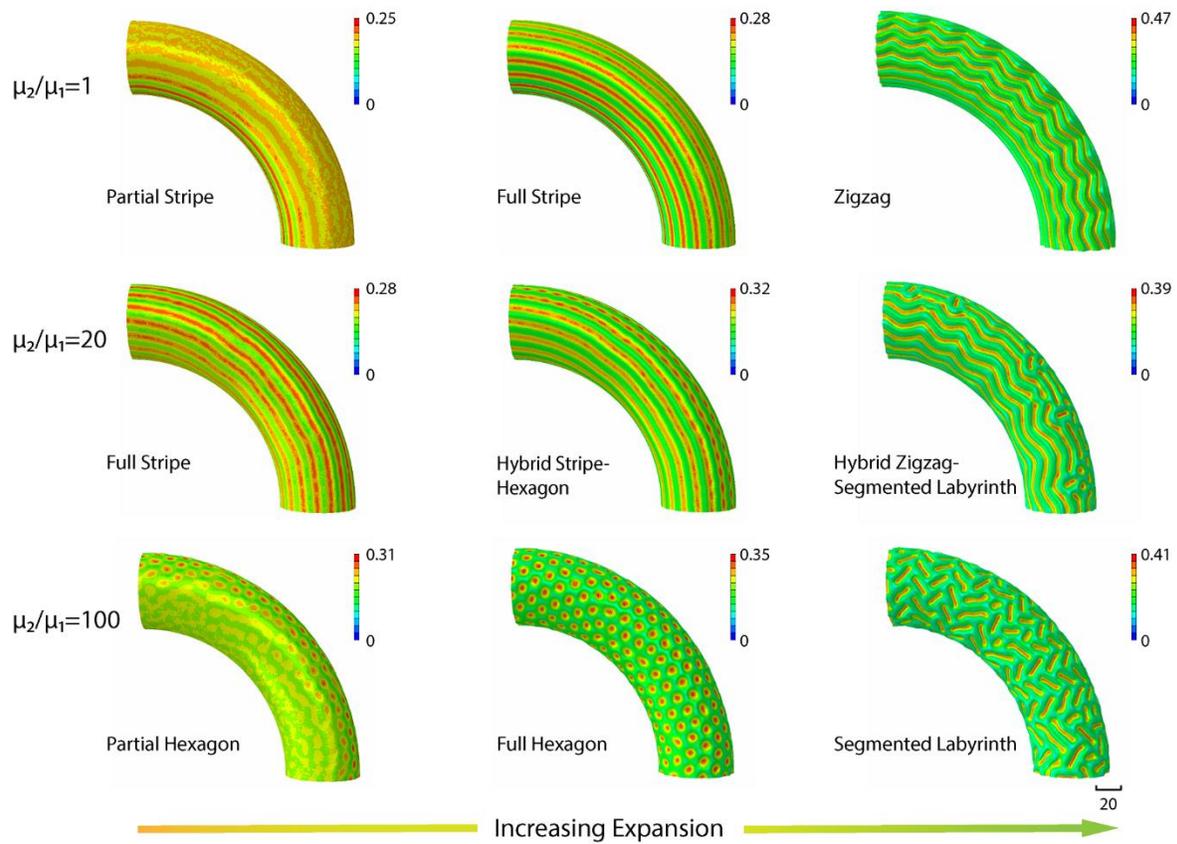

**Fig. 2.** The formation and evolution of wrinkling patterns on a quarter of torus as outer film expansion increases. The different surface morphologies, from top to bottom, are displayed for varying elastic ratios $C = \mu_2/\mu_1$. The colors indicate the maximum principal logarithmic strain.

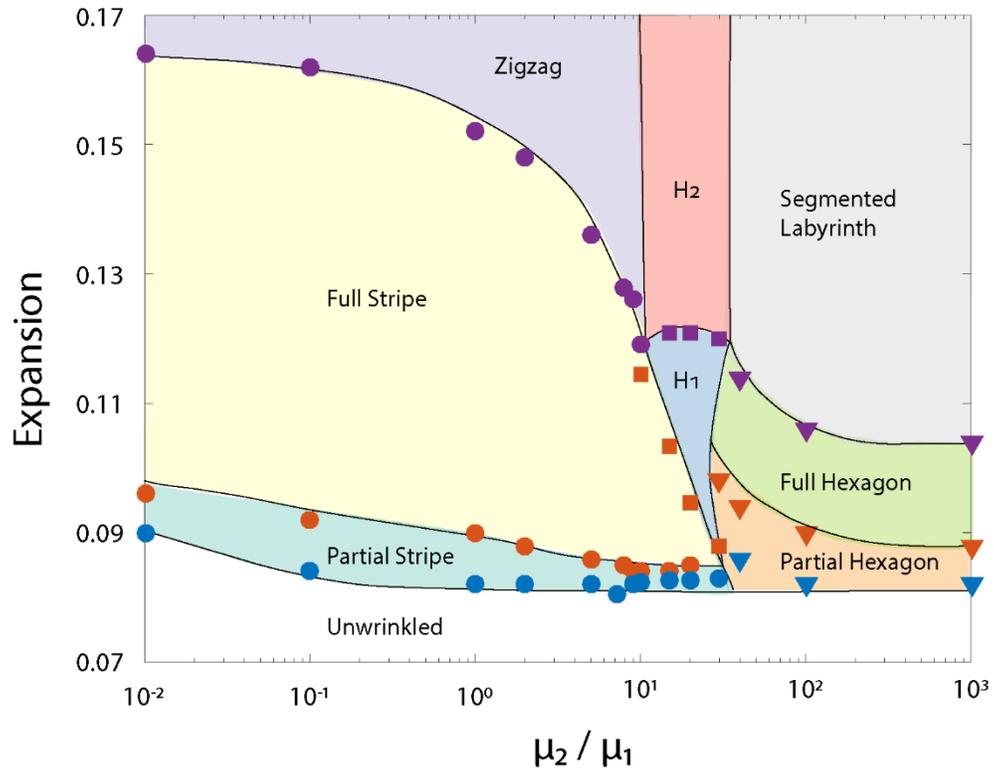

**Fig. 3.** The phase diagram of wrinkling patterns in terms of the elastic ratio and expansion value. H1 and H2 represent the hybrid stripe-hexagon and hybrid zigzag-segmented labyrinth, respectively.

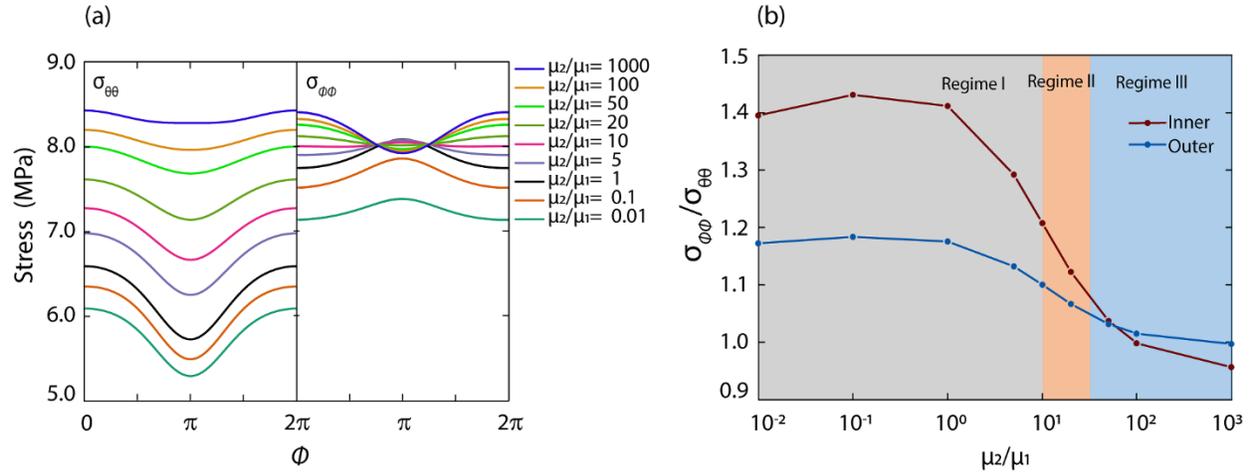

**Fig. 4.** Stresses in the outer film. (a) The absolute value of compressive stresses along toroidal (left, $\sigma_{\theta\theta}$) and poloidal (right, $\sigma_{\Phi\Phi}$) directions. (b) Ratios between poloidal and toroidal stresses at the outer ($\Phi = 0, 2\pi$) and inner ($\Phi = \pi$) surface lines of the torus as a function of elastic ratio $C = \mu_2/\mu_1$.

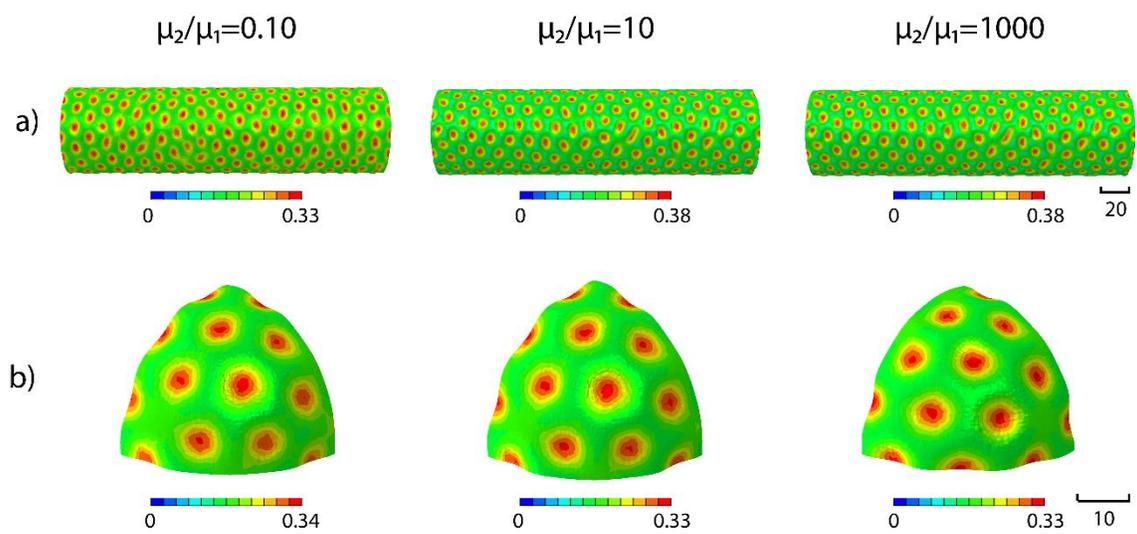

**Fig.5.** Hexagonal patterns on cylinders (a) and one-eighth of a sphere (b) at three different elastic ratios: $C = \mu_2/\mu_1 = 0.10, 10, 1000$. The colors indicate the maximum principal logarithmic strain.

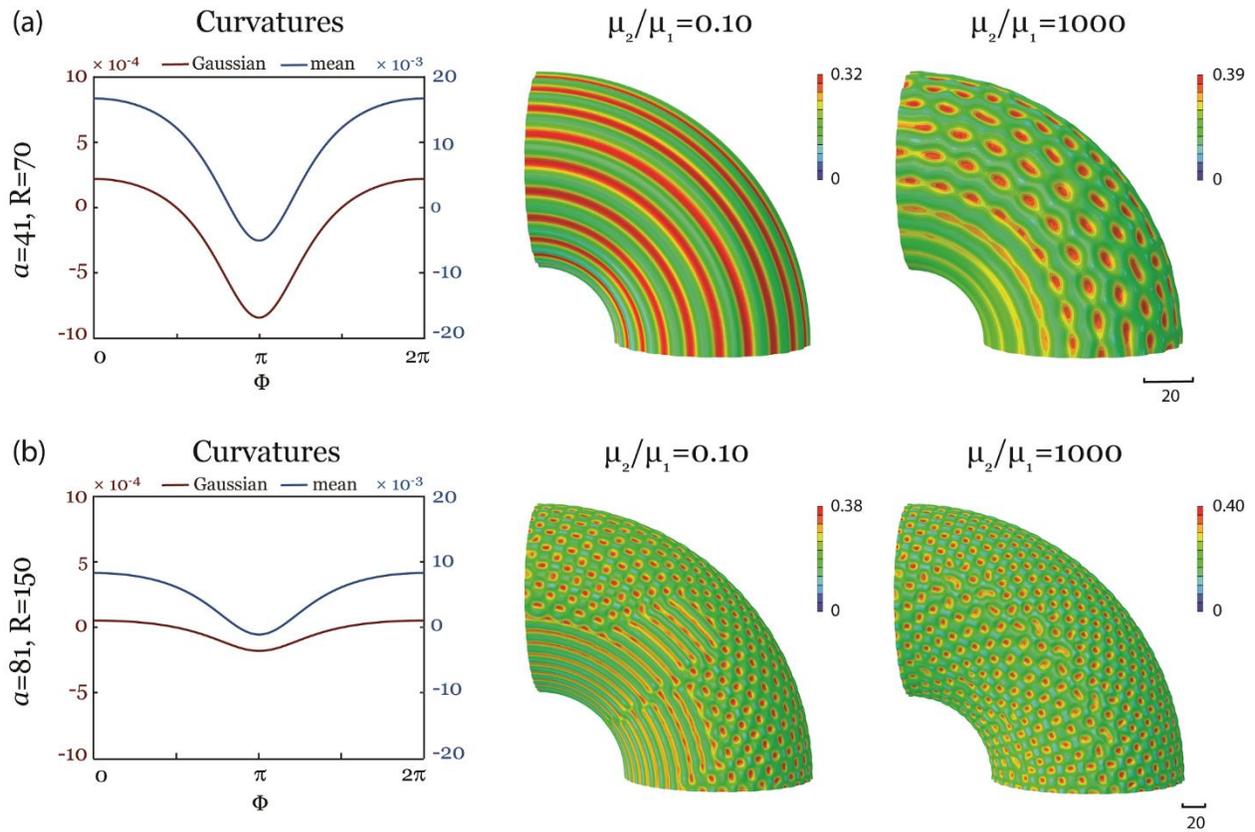

**Fig.6.** Patterns on torus with $a = 41, R = 70$ (a) and $a = 81, R = 150$ (b) at two different elastic ratios: $C = \mu_2/\mu_1 = 0.10, 1000$. The colors indicate the maximum principal logarithmic strain.